\documentstyle[12pt,aps,preprint]{revtex}
\draft
\title{Static Einstein's Universe as a Quantum Solution of Causal Quantum Gravity}
\author{ALI SHOJAI\thanks{Email: FATIMAH@IPM.IR}}
\address{Physics Department, Tehran University, End of North karegar St., Tehran 14352, IRAN}
\address{and}
\address{Institute for Studies in Theoretical Physics and Mathematics, P.O.Box 19395-5531, Tehran,IRAN}
\author{FATIMAH SHOJAI\thanks{Email: SHOJAI@IPM.IR}}
\address{Physics Department, Tehran University, End of North karegar St., Tehran 14352, IRAN}
\address{and}
\address{Institute for Studies in Theoretical Physics and Mathematics, P.O.Box 19395-5531, Tehran,
IRAN}
\author{NARESH DADHICH\thanks{Email:NKD@IUCAA.ERNET.IN}}
\address{Inter-University Centre for Astronomy \& Astrophysics,
Post Bag 4, Pune~411~007, India}
\begin{document}
\maketitle
\begin{abstract}
We shall in the framework of Bohmian quantum gravity show that it is possible to find a {\it pure\/} quantum state which leads to the static Einstein universe whose classical counterpart is flat space--time. We obtain the solution not only in the long--wavelength approximation but also exactly.
\end{abstract}
\pacs{PACS NO.: 03.65.Ta; 04.60.Ds}
\section{Introduction}
In a series of papers\cite{salopek}, a new systematic method is presented for solving the Hamilton--Jacobi equation of general relativity with matter fields. This method is based on expanding Hamilton--Jacobi functional in terms of the spatial gradients of the fields. Such a functional must be 3--diffeomorphism invariant at each order of expansion, in order that the momentum constraint is satisfied automatically. This method is valid only when the scale of the spatial variations of the fields is larger than the characteristic scale of the system. The important aspect of this method is that it can lead to the long--wavelength perturbative solutions without imposing any spatial symmetry. 

This method is extended to Brans--Dicke theory\cite{soda}, $n$--dimensional gravity \cite{chiba} and gravity interacting with electromagnetic field \cite{darian}. Applying this approximation to the low energy effective string action\cite{say} gives a quasi--isotropic solution of pre-Big Bang cosmology and some results about duality transformation. At the quantum level, a similar perturbative method is presented \cite{shojai} to solve the quantum Hamilton--Jacobi equation in de-Broglie--Bohm approach to quantum gravity. This explicitly gives a perturbative solution of quantum gravity. As an application, the authors have investigated the pure quantum solutions of Bohmian quantum gravity \cite{pure} in which quantum potential cannot be ignored with respect to classical potential. Since in the quantum gravity the characteristic length is the Plank  length, it is obvious that the spatial variation scale of the fields have to be larger than it for  having a converging expansion.  

In this paper we shall lay emphasis on applications of the long--wavelength approximation in quantum gravity without matter. In the next section we shall give an overview of Bohmian quantum gravity which will be followed by the solution of the regularized quantum Hamilton--Jacobi--Einstein (HJE) equation in the minisuperspace of spatially conformally flat spaces. We conclude with a discussion.  

\section{de-Broglie--Bohm quantum gravity}
Given any quantum system, there is a general procedure to interpret it causally. This procedure is usually called the de-Broglie--Bohm interpretation of quantum mechanics. According to de-Broglie--Bohm theory, one uses the similarity between the Hamilton--Jacobi theory and quantum mechanics. To explain what we mean, let us start from the non--relativistic quantum mechanics. If one uses $\psi=|\psi|\exp(iS/\hbar)$ in the Schr\"odinger equation and separate the real and imaginary parts we get:
\begin{equation}
\frac{\partial S}{\partial t}+\frac{|\vec{\nabla}S|^2}{2m}+V+Q=0,
\end{equation}
\begin{equation}
\frac{\partial |\psi|^2}{\partial t}+\vec{\nabla}\cdot\left ( |\psi|^2\frac{\vec{\nabla}S}{m}\right )=0.
\end{equation}
The first equation is a Hamilton--Jacobi like equation. The only difference is the presence of quantum potential defined as:
\begin{equation}
Q=-\frac{\hbar^2}{2m}\frac{\nabla^2|\psi|}{|\psi|}.
\end{equation}
The second equation is the continuity equation for an ensemble of the system under consideration. This similarity with the classical mechanics suggests that in order to get causal trajectories, one must use the momentum definition as a guiding relation:
\begin{equation}
\vec{p}=m\vec{v}=\vec{\nabla}S.
\end{equation} 
It is necessary to show that this leads to a coherent description of quantum mechanics. For example, one must show that there is no contradiction with the measurement axiom of quantum mechanics or with the uncertainty principle. The issue of self--consistency of the causal interpretation is well studied in the literature and it is shown that this indeed gives a coherent picture \cite{holland}.

The importance of the causal description becomes more clear when one uses it for systems like general relativity. Since in the causal picture the Hilbert space structure has a secondary role, standard problems of quantum gravity related to the Hilbert space structure may be handled more easily. In addition, conceptual problems like the problem of time would not exist. 

Starting from the standard Hamiltonian formulation of general relativity in terms of the ADM decomposition, after quantization one gets the WDW equation (setting $16\pi G=1$):
\begin{equation}
\hbar^2 h^{-q}\frac{\delta}{\delta h_{ij}}\left ( h^qG_{ijkl}\frac{\delta \Psi}{\delta h_{kl}}\right ) +\sqrt{h} {\cal R}\Psi+\frac{1}{\sqrt{h}}{\cal T}^{00}(-i\hbar\delta/\delta\phi_a,\phi_a)\Psi =0
\end{equation}
and the 3--diffeomorphism constraint:
\begin{equation}
2\nabla_j\frac{\delta}{\delta h_{ij}}\Psi-{\cal T}^{i0}(\delta/\delta\phi_a,\phi_a)\Psi=0
\end{equation}
where $h_{ij}$ is the spatial 3--metric and $h$ and ${\cal R}$ are its determinant and 3-curvature scalar respectively. $q$ is the ordering parameter, and $G_{ijkl}=\frac{1}{\sqrt{h}}(h_{ik}h_{jl}+h_{il}h_{jk}-h_{ij}h_{kl})$ is the superspace metric. ${\cal T}^{\mu\nu}$ is the energy momentum tensor of matter fields ($\phi_a$) in which the matter is quantized by replacing its conjugate momenta by $-i\hbar\delta/\delta\phi_a$. 

In order to get the causal interpretation one should set $\Psi=|\Psi|\exp(iS/\hbar)$ and the resulting equations are:
\begin{equation}
G_{ijkl}\frac{\delta S}{\delta h_{ij}}\frac{\delta S}{\delta h_{kl}}-\sqrt{h}({\cal R}-Q_G)+\frac{1}{\sqrt{h}}\left ( {\cal T}^{00}(\delta S/\delta\phi_a,\phi_a)+Q_M\right ) =0,
\label{a}
\end{equation}
\begin{equation}
\frac{\delta}{\delta h_{ij}}\left ( 2h^qG_{ijkl}\frac{\delta S}{\delta h_{kl}}|\Psi|^2\right ) +\sum_a \frac{\delta}{\delta\phi_a}\left ( h^{q-1/2}\frac{\delta S}{\delta \phi_a}|\Psi|^2\right )=0,
\end{equation}
\begin{equation}
Q_G=-\frac{\hbar^2}{\sqrt{h}|\Psi|}\left ( h^{-q}\frac{\delta}{\delta h_{ij}} h^q G_{ijkl}\frac{\delta|\Psi|}{\delta h_{kl}}\right ),
\end{equation}
\begin{equation}
Q_M=-\frac{\hbar^2}{h|\Psi|}\sum_a \frac{\delta^2|\Psi|}{\delta\phi_a^2},
\end{equation}
\begin{equation}
2\nabla_j\frac{\delta|\Psi|}{\delta h_{ij}}-{\cal T}^{i0}(\delta|\Psi|/\delta\phi_a,\phi_a)=0,
\end{equation}
\begin{equation}
2\nabla_j\frac{\delta S}{\delta h_{ij}}-{\cal T}^{i0}(\delta S/\delta\phi_a,\phi_a)=0.
\label{b}
\end{equation}
It must be noted here that in the above equations all terms containing the second functional derivative are ill--defined and hence they must be regularized. We do this by changing terms like $\delta/\delta h_{ij}(x)\delta/\delta h_{ij}(x)$ into $\int d^3x' \sqrt{h} U(x-x') \delta/\delta h_{ij}(x) \delta/\delta h_{ij}(x')$, where $U$ is the regulator.

The last two equations are conditions on the norm and phase of the wave functional to be diffeomorphism invariant. The first two ones are the quantum Hamilton--Jacobi--Einstein equation and the continuity equation. The guiding relations are:
\begin{equation}
\pi^{kl}=\sqrt{h}(K^{kl}-Kh^{kl})=\frac{\delta S}{\delta h_{kl}}
\label{g}
\end{equation}
\begin{equation}
\pi_{\phi_a}=\frac{\delta S}{\delta \phi_a}
\end{equation}
where $K_{ij}=\frac{1}{2N}(\dot{h}_{ij}-\nabla_iN_j-\nabla_jN_i)$ is the extrinsic curvature and $N$ and $N_i$ are the lapse and shift functions.

One can either solve the WDW equation or solve the relations (\ref{a})--(\ref{b}) in order to get the causal trajectories. Using the quantum Hamilton--Jacobi--Einstein equation, one can define a limit which is called the pure quantum limit. In this limit the total quantum potential is of the same order as the total classical potential. Then it is possible for them to cancel each other, which is the case we are interested in here.  In this case one has:
\begin{equation}
\frac{\delta S}{\delta h_{ij}}=\frac{\delta S}{\delta \phi_a}=0.
\label{qw}
\end{equation}  

This situation is completely opposite to the classical limit in which the quantum potential can be ignored with respect to the classical one. Therefore in the pure quantum limit one gets trajectory of metric, which is not similar to any classical solution. Also as $\frac{\delta S}{\delta h_{kl}}=0$, according to (\ref{g}), the gravitational system would be static. 

The important point here is that because of the relation (\ref{qw}), the continuity equation would be satisfied identically. Also the quantum HJE equation for a pure quantum state is in fact an equation for spatial dependence of the metric and matter fields in terms of any given norm of the wave function.
\section{Solution of quantum Hamilton --Jacobi --Einstein equation for conformally spatially flat mini super space}
According to the previous section, in the causal canonical approach to quantum gravity the probability amplitude and the quantum Hamilton--Jacobi functional satisfy the following four equations:
\begin{equation}
G_{ijkl}\frac{\delta S}{\delta h_{ij}}\frac{\delta S}{\delta h_{kl}}-\sqrt{h}({\cal R}-Q_G)=0,
\label{a1}
\end{equation}
\begin{equation}
\frac{\delta}{\delta h_{ij}}\left ( 2h^qG_{ijkl}\frac{\delta S}{\delta h_{kl}}|\Psi|^2\right ) =0,
\end{equation}
\begin{equation}
2\nabla_j\frac{\delta|\Psi|}{\delta h_{ij}}=0,
\end{equation}
\begin{equation}
2\nabla_j\frac{\delta S}{\delta h_{ij}}=0,
\label{b1}
\end{equation}
where for simplicity we have ignored the matter fields. These equations show how the general covariance of general relativity appear after 3+1 decomposition (ADM) at the quantum level. As usual one can obtain some solutions of these equations by employing mini superspace. In order to get the three diffeomorphism constraints automatically, one must write all functionals in an apparent three scalar form. Then only we can address the first two equations. Moreover as
we are dealing with the pure quantum case, the continuity equation is automatically  satisfied. Therefore it is sufficient to consider the quantum HJE equation. 

Let us write the space--time metric in the following form:
\begin{equation}
ds^2=-N^2dt^2+\exp(2\Lambda)f_{ij}dx^idx^j
\end{equation}
where $N$ is the lapse function, $f_{ij}$ is a pre--prescribed flat 3-metric and $\Lambda$ is the dynamical degree of freedom and for simplicity we choose a gauge in which the shift function is zero. It is clear that since we have no matter field at all, a classical trivial solution is the flat space--time. The 3-curvature of the above metric is:
\begin{equation}
{\cal R}=\exp(-2\Lambda)(4\nabla^2\Lambda+2|\vec{\nabla}\Lambda|^2)
\label{r3}
\end{equation}
and the regularized quantum HJE equation would read as
\[
\int d^3x'\sqrt{f}U(x-x')\left \{ \ell_p^{-3}\frac{\delta^2\Omega}{\delta\Lambda(x)\delta\Lambda(x')}
+\ell_p^{-6}\frac{\delta\Omega}{\delta\Lambda(x)}\frac{\delta\Omega}{\delta\Lambda(x')} \right .
\]
\begin{equation}
\left . +(6q-1)\ell_p^{-3} \frac{\delta\Omega}{\delta\Lambda(x)}\delta(x-x') \right \} = \frac{48f(x)}{\hbar^2}\exp(4\Lambda(x))(2\nabla^2\Lambda+|\vec{\nabla}\Lambda|^2)
\label{x}
\end{equation}
where $\Omega=\ell_p^3\ln |\Psi|$ and $f=\det(f_{ij})$. 

\subsection{Long--wavelength solution}
Now we shall apply the perturbative method of \cite{salopek,shojai} to solve the WDW equation. We shall restrict  up to the second order, but in principle one can go to any order.
Expanding the probability functional in terms of the spatial gradients of the conformal factor:
\begin{equation}
\Omega=\sum_{n=0}^\infty \Omega^{(2n)}.
\end{equation}
\subsubsection{Zeroth order solution}
We write
\begin{equation}
\Omega^{(0)}=\int d^3z \sqrt{f(z)} F[\Lambda]
\end{equation}
which is obviously invariant under spatial coordinate transformations and contains no spatial gradients. Substituting this in the quantum HJE equation (\ref{x}), we have
\begin{equation}
\frac{d^2F}{d\Lambda^2}+\left ( \frac{dF}{d\Lambda}\right )^2+(6q-1)\frac{dF}{d\Lambda}=0.
\end{equation}  
Note that in equation (\ref{x}), in the second term the regulator does nothing. This term needs no regulator, so the regulator can be changed with a sharp function as: $\sqrt{f}U(x-x')=\ell_p^3U(0)\delta (x-x')$.

The solution to this equation is:  
\begin{equation}
F[\Lambda]=B + \alpha \Lambda + \ln (1-A \exp(-\alpha\Lambda))
\end{equation}
where $\alpha=1-6q$ and  $A$ and $B$ are integration constants. 
\subsubsection{Second order solution}
The most general form of the second order term is:
\begin{equation}
\Omega^{(2)}=\int d^3z \sqrt{f(z)} G[\Lambda]|\vec{\nabla}\Lambda|^2.
\end{equation}
Note that the terms like $\nabla ^2 \Lambda$ can be transformed back to $|\vec{\nabla}\Lambda|^2$ by integration by parts. This shows that ${\cal R}$ given by equation (\ref{r3}) is also included in the second order. After substituting this in the equation (\ref{x}) and using the zeroth order solution, we get the following equation:
\[
|\vec{\nabla}\Lambda|^2\left ( G''+\frac{\alpha AG'e^{-\alpha\Lambda}}{1-Ae^{-\alpha\Lambda}}\right )+2\nabla^2\Lambda\left ( G'+ \frac{\alpha AGe^{-\alpha\Lambda}}{1-Ae^{-\alpha\Lambda}} \right )
\]
\begin{equation}
+\frac{48e^{4\Lambda}}{\hbar^2}(2\nabla^2\Lambda+|\vec{\nabla}\Lambda|^2)=0.
\label{er}
\end{equation}  
For simplicity from now on we assume $A=0$. 
As we discussed earlier, according to quantum HJE equation in pure quantum state corresponding to any norm of wave functional there is some spatial dependence of the conformal factor. So one can consider the equation (\ref{er}) as a partial differential equation for $G$ assuming that the spatial derivatives of the conformal factor are given in terms of the conformal factor itself. This means that choosing an arbitrary spatial conformal factor leads to the related wave functional.

Since in the pure quantum case, the metric trajectory is static, a suitable choice of the quantum universe is the static FRW universe given by the space-time metric, 
\begin{equation}
ds^2=-N^2dt^2+\frac{1}{(1+kr^2/4)^2}\left ( dr^2+r^2d\Omega^2\right )
\end{equation}
where $k$ is the curvature parameter. So we have
\begin{equation}
|\vec{\nabla}\Lambda|^2=ke^\Lambda(1-e^\Lambda)
\end{equation}
\begin{equation}
\nabla^2\Lambda=-\frac{k}{2}e^\Lambda \left ( 1+2e^\Lambda \right )
\end{equation}
and substituting in the equation (\ref{er}), we obtain
\begin{equation}
\left ( 1-e^\Lambda\right )G''+\left ( 1+2e^\Lambda\right )G'=-\frac{144}{\hbar^2}e^{5\Lambda}
\end{equation} 
which solves to give 
\[
G=5\ln(e^\Lambda-1)\left ( 6e^\Lambda-2e^{-\Lambda}-e^{2\Lambda}-6\Lambda-3 \right )
-\frac{1}{8}e^{4\Lambda}-\frac{5}{6}e^{3\Lambda}+\frac{1}{2}(\frac{31}{2}-C)e^{2\Lambda}
\]
\begin{equation}
+(-\frac{53}{2}+3C)e^{\Lambda}-(\frac{47}{6}+C)e^{-\Lambda}-(\frac{47}{2}+3C)\Lambda-30 dilog (e^\Lambda)+D
\end{equation}
where $C$ and $D$ are integration constants, and 
\begin{equation}
dilog (x)=\int^x_1 dt \frac{\ln t}{1-t}.
\end{equation}
Thus in this way the wave functional is obtained up to the second order of expansion.
\subsection{Exact solution}
Till now we expand the wave function with respect to the spatial gradients, but there is an exact solution of equation (\ref{x})assuming that the wave function contains no spatial gradients. Setting $\Omega=\int d^3z \sqrt{f} F[\Lambda]$, 
the equation (\ref{x}) would become
\begin{equation}
F''+F'^2-\alpha F'=-\frac{144k}{\hbar^2}e^{6\Lambda}.
\end{equation}
with the solution:
\begin{equation}
F=\ln\left (E_1 J_{|\alpha|/2}(\frac{4\sqrt{k}}{\hbar}e^{3\Lambda})+ E_2 Y_{|\alpha|/2}(\frac{4\sqrt{k}}{\hbar}e^{3\Lambda})\right )
\end{equation}
where $E_1$ and $E_2$ are integration constants.
\section{Concluding remarks}
We have here shown that the static Einstein universe can be thought as a pure quantum universe. That is, starting from a flat space--time, the quantum fluctuations are able to give rise to a causal trajectory for space-time metric which is exactly the Einstein universe. It is interesting to note that the Einstein 
universe can be obtained as a long wavelength perturbative solution as well as an exact solution. This fact is perhaps indicative of stability of the solution.

Another point to note is that our solution is not a perturbation of the classical solution as it is a pure quantum state trajectory and admits no classical limit. The next generalization would be to include matter fields. That would require keeping allowed spatial gradients of matter fields in the expansion terms. 

\end{document}